\documentclass[a4paper]{article}

\usepackage{geometry}

\usepackage{siunitx}
\usepackage[utf8]{inputenc}
\usepackage[T1]{fontenc}
\usepackage{commath,mathtools}\usepackage{dsfont}

\usepackage{tikz}
\usetikzlibrary{plotmarks,calc,math,shapes}

\usepackage{pgfplots}
\usepackage{pgfplotstable}
\usepackage{colortbl}
\usepgfplotslibrary{colorbrewer}

\usepackage{todonotes}

\usepackage{graphicx}

\usepackage{booktabs}
\usepackage{subcaption}
\usepackage[colorlinks=true,citecolor=blue]{hyperref}

\usepackage[backend=bibtex,sorting=none,citestyle=numeric-comp,maxbibnames=9]{biblatex}  
\newtheorem{remark}{Remark}
\usepackage{xspace}

\newcommand{\pf}{PF\xspace}

\newcommand{\buses}{\mathcal{N}}
\newcommand{\lines}{\mathcal{L}}
\newcommand{\Ybus}{Y}
\newcommand{\Gbus}{G}
\newcommand{\Bbus}{B}

\newcommand{\ngeneric}{n}
\newcommand{\nbus}{\ngeneric_{\text{bus}}}
\newcommand{\nline}{\ngeneric_{\text{line}}}

\newcommand{\activepower}{P}
\newcommand{\reactivepower}{Q}

\newcommand{\vmagnitude}{V}
\newcommand{\vangle}{\varphi}

\newcommand{\PQbus}{PQ\xspace}
\newcommand{\PVbus}{PV\xspace}

\DeclareMathOperator{\sign}{sign}

\makeindex             

\begin{document}
\date{\today}
\title{Modeling and simulation of sector-coupled networks:
A gas-power benchmark}
\author{E. Fokken\footnote{E. Fokken,  Department of Mathematics, University of Mannheim. \href{mailto:fokken@uni-mannheim.de}{fokken@uni-mannheim.de}}, \; T. Mühlpfordt\footnote{T. Mühlpfordt,  Institute for Automation and Applied Informatics, Karlsruhe Institute of Technology. \href{mailto:tillmann.muehlpfordt@kit.edu}{tillmann.muehlpfordt@kit.edu}}, \; T. Faulwasser\footnote{T. Faulwasser,  Institute for Energy Systems, Energy Efficiency and Energy Economics, TU Dortmund University. \href{mailto:timm.faulwasser@ieee.org}{timm.faulwasser@ieee.org}}, \; S. G\"ottlich\footnote{S. G\"ottlich,  Department of Mathematics, University of Mannheim. \href{mailto:goettlich@uni-mannheim.de}{goettlich@uni-mannheim.de}}, \; O. Kolb\footnote{O. Kolb,  Department of Mathematics, University of Mannheim. \href{mailto:kolb@uni-mannheim.de}{kolb@uni-mannheim.de}}}

\maketitle

\abstract{In this contribution, we aim at presenting a gas-to-power benchmark problem that can be used for the simulation of electricity and gas networks in a 
time-dependent environment.
Based on realistic data from the IEEE database and the GasLib suite, we describe the full set up of the underlying equations and motivate the choice of parameters.
The simulation results demonstrate the applicability of the proposed approach and also allow for a clear visualization of gas-power conversion.}

\section{Introduction}
\label{sec:introduction}
The current transformation of the energy system is driven by at least three trends: decarbonization and defossilization
of energy supply, increasing concerns about climate change, and political decisions, e.g. the phase-out of nuclear energy \cite{GesetzAtom2011} and foreseen phase-out of coal/lignite in Germany.
Phase out of all fossile energies is in principle doable as averaged over long time spans renewable sources (wind, solar, etc.) provide sufficient amounts of energy to achieve decarbonization.
Yet, renewable generation and demand are not synchronized in time and space and thus energy storage  and transport are both of crucial importance.
Currently, neither a readily and widely usable  storage technology to buffer the quantities of electrical energy required for decarbonization exists, nor is there scientific consensus  on which large-scale storage technologies will be available in near-to-mid future.
Hence, it comes at no surprise that the coupling of energy domains and sectors is gaining increasing research attention.
For example, the economic viability of future power-to-$X$ pathways (where $X$ can be Hydrogen, Methane, or synthetic bio-fuels, for instance) has been investigated in several studies, see e.g.~\cite{Schiebahn15,Brown18a,Heinen16}.
These investigations are driven by the fact that natural gas can be stored in sufficient quantities in dedicated installations and to a certain extend directly in the gas grid itself.
In other words, coupling of electricity and gas networks is currently considered a very promising road towards a high share of renewables. 

Historically, however, the critical energy system infrastructure for gas and power grids has been separated in terms of operation and control.
Hence, there do not exist established standards for joint operation and control of multi-energy grids.
In turn multi-energy systems arising from sector-coupling pose control and optimization challenges, many of which are still open or are subject to ongoing research efforts,~see~\cite{CHERTKOV2015541,ZengFangLiChen2016,Zlotnik16a}, or more recently~\cite{aleksey_michael,2019arXiv190100522F,Fokken2019,OMalley20}. 

In the context of electricity grids, the Optimal Power Flow (OPF) problem is of vital importance for safe, reliable, and economic operation, see~\cite{Frank16,Capitanescu16a,kit:faulwasser18d} for tutorial introductions and overviews.
In its plain form, OPF means to solve a non-convex Nonlinear Program (NLP) of finite  dimension.
This problem can scale up easily to several thousand nodes, hence several thousand decision variables ($4\cdot$ number of nodes).
Among the challenges arising are distributed optimization~\cite{Engelmann2019,Molzahn17a,Kim00a}, convex relaxations~\cite{Low14a,Low14b}, and the consideration of stochastic disturbances~\cite{kit:muehlpfordt18d,Bienstock14a}. 

In principle, similar to electricity grids, gas grids are typically described by a coupled system of 
conservation laws, see e.g.~\cite{coupling_conditions_isothermal,banda_herty_klar,bressan_canic_garavello_herty_piccoli,brouwer_gasser_herty,doi:10.1137/060665841} for an overview. 
However, due to the compressibility of gas, the arising models differ as one has to consider nonlinear and time-dependent 
hyperbolic partial differential equations (PDEs) for the gas flow through pipelines supplemented with appropriate coupling conditions at intersections.
Since these dynamics allow for discontinuous solutions, a careful theoretical and numerical treatment is needed to master the challenges of simulation, optimization, and control for complex networks.
Hence, in the context of gas grids, already the deterministic case is numerically challenging. 
In the context of the so-called energy transition, there is a strong need to couple the different energy supply systems, in particular power-to-gas,
to compensate the differences between supply and demand in the electricity system,
see~\cite{CHERTKOV2015541,ZengFangLiChen2016,Zlotnik2016}, or more recently~\cite{aleksey_michael,2019arXiv190100522F,Fokken2019}.
The resulting mathematical problem requires the design of an appropriate numerical solution method since the transport of electricity happens on a shorter time scale compared to gas.

It is worth to be remarked that for electrical grids and the corresponding optimization problems, there exists a large number of established benchmarks.
This includes IEEE test cases, CIGRE benchmarks~\cite{Strunz06} and, in case of the German system, the scigrid model~\cite{Matke17a}. Likewise for gas networks, the GasLib suite\footnote{\url{http://gaslib.zib.de/}} offers test cases for simulation and optimization purposes, respectively.

However, when it comes to coupled gas-power systems, there do not exist widely accepted benchmarks. One of the few exceptions appears to be the case study presented in \cite{Zlotnik16a}, which comprises the IEEE RTS96 One Area 24 node electrical grid and a 24 pipeline gas network. 
Hence, the present chapter takes further steps towards a more realistic optimization benchmark for multi-energy systems.
Specifically, we couple a model of the Greek gas network\,---\,the gaslib-134 model \cite{Humpola_et_al:2015} which includes 86 pipelines\,---\,with the IEEE 300-bus system under AC conditions. We formulate a combined simulation framework, wherein the key actions are taken for the gas network 
and the electrical side translates into the solution of the AC power flow equations. 
In particular, we observe a significant pressure drop for the gas-plant nodes while the gas consumed varies over time.

\section{Model and algorithm}
\label{sec:model-algorithm-1}
We model the combined power-and-gas network over a time horizon $t \in [0, T]$ by a graph whose nodes and edges represent certain components of the respective networks.
A small example of such a graph is given in
\begin{figure}[h]
  \centering
  \includegraphics[width=0.5\textwidth]{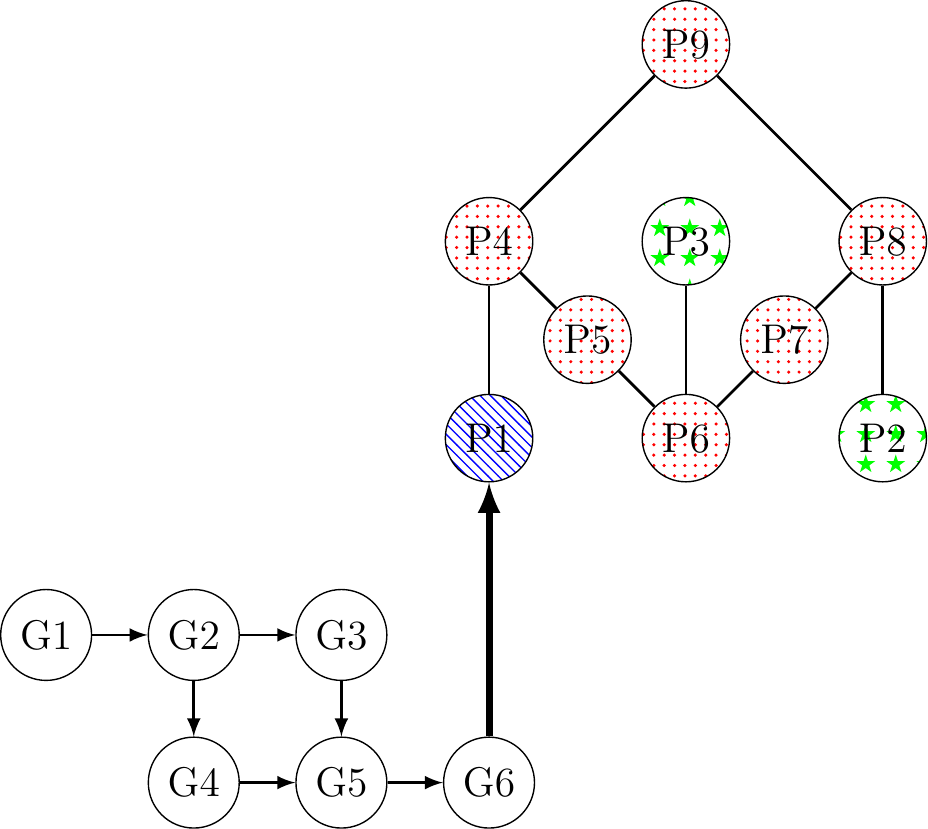}
  \caption{A graph of a combined gas and power network.  In the lower part is a small gas network, the bold arrow represents a gas-power-conversion plant, the upper network consists of different power components.}
  \label{fig:graphexample}
\end{figure}
Every component contributes variables, parameters and/or model equations, see \autoref{sec:gas-model} for the gas model, and \autoref{sec:power-model} for the power model.
We model the coupling between the power grid and the gas network via gas-power-conversion (i.e. a gas power) plants, which are represented by certain edges of the graph. 
Starting from a continuous-time formulation we then discretize the time interval $[0, T]$ to time steps $t_k = k \tfrac{T}{N}$, hence formulating the model equations as an algebraic system of nonlinear equations, which is solved with Newton's method.
The following two sections describe the gas and power model, and \autoref{sec:gas-power-conversion} describes the coupling of both.

\subsection{Model of the gas network}
\label{sec:gas-model}
For the gas network we use the definitions of gaslib~\cite{Humpola_et_al:2015}.  The following theory is also presented in \cite{Fokken2019}.
In our application, a gas edge is one of the following: a pipeline, a so-called short pipe, a (controlled) valve, a compressor station, or a gas-power-conversion plant, where both valves and compressors act as short pipes in our setting.
The nodes represent sources (where gas is injected into the network), sinks (where it is withdrawn) and inner nodes (where no gas is injected or withdrawn).

\subsubsection{Pipelines}
\label{sec:pipelines}
The gas flow in pipelines is modeled by the isentropic Euler equations
\begin{equation}\label{eq:gas}
  \begin{pmatrix}\rho_{l}\\  q_m\end{pmatrix}_t + \begin{pmatrix} q_m\\ p(\rho_{l}) + \frac{ q_m^2}{\rho_{l}}\end{pmatrix}_x = \begin{pmatrix}0\\S(\rho_{l}, q_m) \end{pmatrix},
\end{equation}
where $\rho_l$ is the line density~($\si{\kilogram \per \meter}$) of the gas, $q_m$ is its mass flow ($\si{\kilogram \per \second}$) and~$p$ is the pressure function.
As is often done we employ the (possibly space-dependent) density and the volumetric flow are instead
\begin{equation}
  \label{eq:18}
  \rho =\frac{\rho_l}{A}\ , \ q = \frac{q_m}{\rho_0} \ ,
\end{equation}
where~$A$ is the cross section of the pipe and~$\rho_0$ is the density at standard conditions.
This is relevant for the coupling of pipes with possibly different cross sections, see \autoref{sec:coupling-conditions}.
The subscript indices indicate the partial derivatives where $t \in [0,T]$ is time, and~$x \in [0,L]$ is the position along the pipe of length $L$; $S$ is a friction term given by
\begin{equation}
  S(\rho,q) = -\frac{\lambda(q)}{2d_\text{pipe}} \frac{q \vert q\vert}{\rho}\label{eq:19}.
\end{equation}
The friction factor $\lambda(q)$ is defined by the Prandtl-Colebrook formula \begin{equation}
  \frac{1}{\sqrt{\lambda}} = -2\log_{10} \left( \frac{2.51}{\text{Re}(q) \sqrt{\lambda}} + \frac{k}{3.71 d_\text{pipe}} \right)
\end{equation}
with Reynolds number
\begin{equation}
  \text{Re}(q) = \frac{d_\text{pipe}}{\eta} q\ ,
\end{equation}
roughness of the pipes $k = \SI{8e-6}{\metre}$ and the dynamic viscosity $\eta = \SI{e-5}{\kilogram \per \meter \per \second}$ and the pipe diameter $d_\text{pipe}$.

We use the isothermal pressure function with compressibility factor
\begin{equation}
  \label{eq:8}
  p(\rho) = \frac{c_{\text{vac}}^2\rho}{1-\alpha c_{\text{vac}}^2 \rho}.
\end{equation}
The pressure function can be inverted, yielding
\begin{equation}
  \frac{p}{c_{\text{vac}}^2 z(p)} = \rho, \label{eq:12}
\end{equation}
where $c_{\text{vac}}$ is the limit of the speed of sound in the vacuum limit (that is, for $\rho \to 0$) and $z(p)$ is the compressibility factor.
These parameters are given by
\begin{equation}
  \label{eq:3}
  \begin{aligned}
    c_{\text{vac}} & = \sqrt{\frac{p_0}{z_0}  \frac{T}{T_0} \frac{1}{\rho_0}},\\
    z(p) & = 1+ \alpha p.
  \end{aligned}
\end{equation}
The numerical values for parameters $ \rho_0, p_0,  z_0, T_0, T, \alpha$ are listed in \autoref{tab:standardconditions}.
\begin{table}[t]
  \centering
  \caption{Gas net constants for \eqref{eq:3}.}
  \begin{tabular}{llllll}
  	\toprule
    $\rho_0 \, [\si{\kilogram\per\metre\cubed}]$ & $p_0 \, [\si{\bar}]$ & $z_0$ & $T_0 \, [\si{\kelvin}]$  & $T \, [\si{\kelvin}]$& $\alpha \, [\si{\per \bar}]$\\
    \midrule
    0.785 &     1.01325 &     1.005 &     273.15 &     283.15 &     -0.00224 \\
\bottomrule
  \end{tabular}
  \label{tab:standardconditions}
\end{table}
According to \autoref{tab:standardconditions} we have that $\alpha<0$.
We define $\beta = -\alpha$ and obtain for the pressure function~\eqref{eq:8}.
\begin{align}
  p(\rho) = \frac{c_{\text{vac}}^2\rho}{1+\beta c_{\text{vac}}^2\rho} =  \frac{1}{\beta} + \frac{1}{\beta^2c_{\text{vac}}^2}\left(\frac{1}{\rho +\frac{1}{\beta c_{\text{vac}}^2}}\right).  \label{eq:11}
\end{align}

Let us check whether this pressure function is valid for the well-posedness of the isentropic Euler equations.
Note that, according to Proposition C in~\cite{Fokken2019}, validity of a pressure function is unchanged by adding or multiplying a positive constant.
Therefore \eqref{eq:11} is valid if the innermost bracket $\left(\rho +\frac{1}{\beta c_{\text{vac}}^2}\right)^{-1}$ is valid.

As PDEs, the model equations~\eqref{eq:gas} for the pipelines are infinite-dimensional. 
To use them in our algorithm we need to choose a discretization in both time and space.
To this end we employ the implicit box scheme~\cite{KolbLangBales2010}, which is of the form
\begin{equation}
  \label{eq:7}
  \frac{U^{*}_{j-1} + U^{*}_{j}}{2} = \frac{U^{}_{j-1} + U^{}_{j}}{2} - \frac{\Delta t}{\Delta x}\left( F(U^{*}_j)- F(U^{*}_{j-1}) \right) + \Delta t \frac{G(U^{*}_j)+G(U^{*}_{j-1})}{2}\ .
\end{equation}
where $\Delta t = \tfrac{T}{N}$ is the time step-size, $\Delta x = \tfrac{L}{M}$ the space step-size, and
\begin{equation}
  \label{eq:9}
  U_j = \begin{pmatrix}\rho(k \Delta t,j \Delta x)\\q(k \Delta t,j \Delta x) \end{pmatrix} \ \text{and}\ U_j^{*} = \begin{pmatrix}\rho((k+1) \Delta t,j \Delta x)\\q(k \Delta t,j \Delta x) \end{pmatrix} \ ,
\end{equation}
are the states at the last time step and the current time step, respectively.
Variables with superscript $(\cdot)^*$ are the unknowns to be computed.
In the box scheme~\eqref{eq:7}, the flux term~$F$, and the source term~$G$ are given by
\begin{equation}
  \label{eq:16}
  F(U) = \begin{pmatrix} q\\ p(\rho) + \frac{ q^2}{\rho}\end{pmatrix} \ \text{and} \ G(U) = \begin{pmatrix}0\\S(\rho, q) \end{pmatrix}.
\end{equation}
\subsection{Other edges}
\label{sec:other-edges}
All other edges\,---\,apart from the gas-power-conversion plants\,---\,act like a short pipe in our setting.
A short pipe has no physical properties and exactly two states, corresponding to the beginning and the end of it respectively.
Their model equations set these states to be equal.
That is, for a short pipe with incoming state $(\rho_\text{in}, q_\text{in})$ and outgoing state $(\rho_\text{out}, q_\text{out})$ there holds $\rho_\text{in} =\rho_\text{out}$ and $q_\text{in} = q_\text{out}$.
Short pipes can be used to separate boundary conditions from coupling conditions on the computational level by inserting the artificial short pipe in between a node with multiple attached pipes and a source/sink node.

\subsection{Nodes}
\label{sec:coupling-conditions}
The nodes of the gas network comprise the algebraic equations coupling the states of the corresponding edges.
In addition, the source nodes and the sink nodes entail the boundary conditions describing inflow  and outflow, respectively.
We use two kinds of boundary conditions, which share the same structure:
On the one hand, we enforce equality of the pressure at a node.
On the other hand, we demand Bernoulli invariant coupling conditions to be satisfied as introduced in~\cite{reigstad2}.
These coupling conditions yield entropy reduction at the nodes.
Specifically, at a node with $l \in \mathds{N}$ attached edges, the coupling conditions are of the form
\begin{equation}
  \label{eq:2}
  \begin{aligned}
    \sum_{k = 1}^lq_k &=0\\
    H(\rho_k,q_k)&=H(\rho_{k-1},q_{k-1}) \text{ for } k = 2,\dots, l.
      \end{aligned}
\end{equation} 
The pressure coupling condition reads
\begin{equation}
  \label{eq:17}
  H_p(\rho,q) = p(\rho),
\end{equation}
while the Bernoulli coupling condition is
\begin{equation}
  \label{eq:1}
  H_b(\rho,q) = \frac{1}{2}\left(\frac{\rho_0 q}{\rho A}\right)^2 +\int_{\rho_0}^{\rho}\frac{p^{\prime}(\hat{\rho})}{\hat{\rho}}\dif \hat{\rho}.
\end{equation}
Note that we use the space-dependent density, not the line density.\footnote{The line density is usually discontinuous over a node, whereas the three-dimensional space-dependent density is not in case of~$H_p$.}
Because $\rho \mapsto p(\rho)$ is one-to-one \eqref{eq:1} can be written as
\begin{equation}
\label{eq:20}
  H_b(p,q) = \frac{1}{2}\left(\frac{\rho_0 q}{\rho(p) A}\right)^2 +\int_{p_0}^{p}\frac{1}{\rho(\hat{p})}\dif \hat{p}\ .
\end{equation}
Note that $v = \tfrac{\rho_0 q}{\rho A}$ is simply the flow velocity of the gas.
If we were to omit the first part of~$H_b$, this would be equivalent to the usual condition of pressure equality. The integral in \eqref{eq:20} can be solved when~$\rho(p)$ is inserted from~\eqref{eq:12}. This yields
\begin{equation}
  \label{eq:5}
  H_b(p,q) = \frac{1}{2}\left(\frac{\rho_0 q}{\rho(p) A}  \right)^2 +c_{\text{vac}}^2 \left[ \ln\left(\frac{p}{p_0}\right)+\alpha(p-p_0) \right].
\end{equation}
We remark that in our simulation results below, the velocity part of~$H_b$ is almost irrelevant and the coupling constant behaves almost like $H_p$.
This seems plausible in case the velocity is much smaller than the speed of sound, which is true for realistic pipeline settings.

\begin{remark}[Implementing Bernoulli coupling conditions]
  \label{remark:nodes}
  Although $H_b$ represents the better physical model, it brings about implementation issues:
  At a node where in addition to pipelines a short pipe or any other connection type (like a compressor or a valve) is attached, the term~$H_b$ cannot be easily evaluated, as this requires to know the pipe cross section; a quantity often not available for these components.
  A workaround is to set the pressure for short pipes arbitrarily, and only set the Bernoulli invariants for all edges that have a known cross section area.
\end{remark}

\subsection{Model of the power grid}
\label{sec:power-model}
We study a connected electrical grid under steady state conditions in terms of its single-phase equivalent; as is commonly done, we model electrical lines by algebraic relations derived from the so-called $\Pi$-line equivalent~~\cite{Frank16,kit:faulwasser18d,Andersson15book,Grainger94book,Wood2012}.
These standard assumptions simplify the mathematical model of the electrical grid tremendously: instead of partial differential equations for three-phase systems it suffices to study a system of nonlinear algebraic equations, the so-called power flow equations~\cite{Frank16,Andersson15book}.
\begin{figure}
	\centering
	\includegraphics{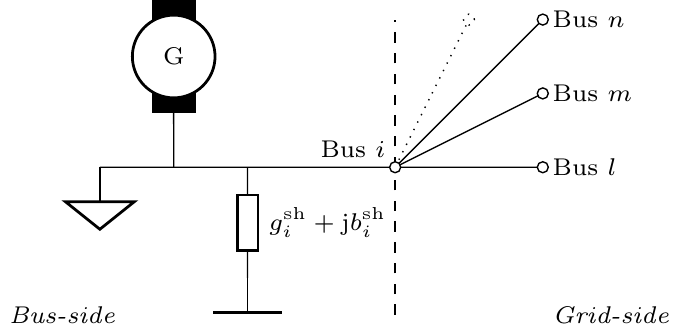}
	\caption{Components at bus $i \in \buses$, and its connection to the remaining grid \cite{MuehlpfordtDissertation}.\label{fig:BusOfPowerSystem}}
\end{figure}

We represent the electrical grid by the triple $(\buses, \lines, \Ybus)$: $\buses = \{ 1, \hdots, \nbus \}$ is the non-empty set of buses, $\lines = \{ 1, \hdots, \nline \}$ is the non-empty set of lines, and $\Ybus = \Gbus + \mathrm{i} \Bbus \in \mathds{C}^{\nbus \times \nbus}$ is the so-called bus-admittance matrix that contains both topological and physical information---such as shunts and line impedances---about the grid \cite{Andersson15book}.\footnote{In graph-theoretic terms, $\buses$ contains the nodes, $\lines$ contains the edges, and $\Ybus$ corresponds to the weighted Graph Laplacian in the absence of phase-shifting transformers. For the construction of the bus admittance matrix~$\Ybus$ there exist explicit formulas, see e.g.~\cite{Andersson15book,Frank16}. Alternatively, power systems software packages such as Matpower~\cite{Zimmerman11} provide the functionality to generate the bus admittance matrix~$\Ybus$ from a given case file (via \texttt{makeYbus(casefile)}).}
Each bus $i \in \buses$ is characterized by its voltage phasor~$\vmagnitude_i \exp( \mathrm{i} \vangle_i)$, and its net apparent power $\activepower_i + \mathrm{i} \reactivepower_i$, with $\mathrm{i}$ being the imaginary unit.
We call $\vmagnitude_i$~voltage magnitude, $\vangle_i$~voltage angle, $\activepower_i$ ~active power, and $\reactivepower_i$~reactive power at bus~$i$, respectively.
The power flow equations---here given in polar form---relate the bus voltages to the powers according to
\begin{subequations}
	\label{eq:PowerFlowEquations}
	\begin{align}
	\activepower_i &= \sum_{j \in \buses} \vmagnitude_i \vmagnitude_j \left( \Gbus_{ij} \cos (\vangle_i - \vangle_j) + \Bbus_{ij} \sin (\vangle_i - \vangle_j) \right), \\
	\reactivepower_i &= \sum_{j \in \buses} \vmagnitude_i \vmagnitude_j \left( \Gbus_{ij} \sin (\vangle_i - \vangle_j) - \Bbus_{ij} \cos (\vangle_i - \vangle_j) \right), 
	\end{align}
\end{subequations}
for all buses~$i \in \buses$.
The \pf equations~\eqref{eq:PowerFlowEquations} constitute $2 \nbus$ nonlinear and non-convex equations in $4 \nbus$ variables.
The remaining $2 \nbus$ degrees of freedoms are fixed by introducing so-called bus specifications, which are listed in \autoref{tab:BusSpecifications}.
A \PQbus represents a load, a \PVbus represents a generator, and the slack bus is a specific generation node that provides an angle reference.
The angle reference is needed to overcome the rotational degeneracy of the \pf equations~\eqref{eq:PowerFlowEquations}.

A power flow study is to solve the nonlinear system of $4 \nbus$ equations built from the $2 \nbus$ \pf equations~\eqref{eq:PowerFlowEquations} together with the $2 \nbus$ bus specifications from \autoref{tab:BusSpecifications}.
We concisely write
\begin{equation}
\label{eq:PowerFlowStudy}
g(x) = 0,
\end{equation}
where $x$ contains the voltage magnitude, the voltage angle, the active power, and the reactive power of every bus~$i \in \buses$.  These are the model equations provided by the power network.
Arguably, there exists a plethora of methods to solve the system~\eqref{eq:PowerFlowStudy}, the Newton method, which is also used by us, being perhaps the most prevalent one~\cite{Frank16,Andersson15book,Grainger94book}.

\begin{table}[]
	\centering
	\caption{Common bus specifications for bus $i \in \buses$ \cite{Frank16,Andersson15book,Grainger94book}.\label{tab:BusSpecifications}}
	\begin{tabular}{llll}
		\toprule
		Bus name & \multicolumn{3}{c}{Fixed quantities} \\
		\midrule
		\PQbus & Active power $\activepower_i$ & \& & reactive power $\reactivepower_i$\\
		\PVbus & Active power $\activepower_i$ & \& & voltage magnitude $\vmagnitude_i$\\
		Slack & Voltage angle $\vangle_i$ & \& & voltage magnitude $\vmagnitude_i$\\
		\bottomrule
	\end{tabular}
\end{table}

\subsection{Gas-Power-Conversion}
\label{sec:gas-power-conversion}
Having covered the modeling of both the gas and the electrical side, how are they connected?
Each gas-power-conversion plant is modeled as an edge between the gas network nodes and the power grid nodes listed in~\autoref{tab:gas-power-interface}.
These operate in two modes, namely Gas-to-Power (GtP), that is as a gas power plant and Power-to-Gas (PtG), where surplus electric power is converted to natural gas, e.g. by electrolysis and methanisation.
The simplified model equations have the same form in both cases, namely
\begin{equation}
  \label{eq:4}
  q = E_\text{mode}(\sign (\activepower)) \activepower,
\end{equation}
where $\activepower$ is the power demand (positive) or supply (negative) of the connected power node, $q$ is the outflow of the sink, and $E_\text{mode}$ a conversion factor of PtG and GtP conversion, respectively.
This piece-wise linear model serves as an approximation of the heat rate of a power plant, respectively the efficiency of a PtG-plant.
To overcome the non-differentiability of~\eqref{eq:4} at $\activepower = 0$, we employ an interpolating function~$S$
\begin{equation}
  \label{eq:13}
  S(x,a,b,\epsilon) = x\left(\frac{1}{2}(a+b)-\frac{3}{4}(b-a)\frac{x}{\epsilon} +\frac{(b-a)}{4}\left(\frac{x}{\epsilon}\right)^3\right),
\end{equation}
with properties
\begin{itemize}
\item $S$ is a polynomial of degree $4$ in $x$,
\item $S(0,a,b,\epsilon) = 0$,
\item $S(\epsilon,a,b,\epsilon) = a \cdot\epsilon$,
\item $S(-\epsilon,a,b,\epsilon) = b\cdot  (-\epsilon)$,
\item $\pd{S}{x}(\epsilon,a,b,\epsilon) = a $,
\item $\pd{S}{x}(-\epsilon,a,b,\epsilon) = b $.
\end{itemize}
Hence, we replace the conversion~\eqref{eq:4} by
\begin{equation}
  \label{eq:15}
  q =
  \begin{cases}
    E_{\text{PtG}}\cdot\activepower &\text{ for } \phantom{ {}- \epsilon < {}} \activepower<-\epsilon\\
    S(P,E_{\text{GtP}},E_{\text{PtG}},\epsilon) &\text{ for } -\epsilon <\activepower<\epsilon \\
    E_{\text{GtP}}\cdot\activepower &\text{ for } \phantom{{}-{}}\epsilon<\activepower 
  \end{cases}\ ,
\end{equation}
which makes $\activepower \mapsto q(\activepower) \in C^1(\mathds{R})$ and $q(0) = 0$, so that no gas is taken from or injected into the gas network if electrical power is neither drawn nor supplied.

\section{Network data}
\label{sec:network-data}

\subsection{Gas network}
\label{sec:gas-network}
We use the gaslib-134 model \cite{Humpola_et_al:2015} with inactive compressor and inactive valve.
This is a network with $90$ sink nodes, $3$ source nodes and $86$ inner nodes.
As connections, there are $86$ pipes, $45$ short pipes, one compressor and one valve.
The pipes have a total length of approximately $1500\si{\kilo\metre}$. As valve and compressor are inactive, they just let gas flow through them.
We let them act like short pipes with the following exception.
Although gaslib-134 doesn't provide one we attach a cross section to these components so they can partake in the Bernoulli coupling \eqref{eq:1}.
The compressor begins at the end of a single pipe and the valve ends at a single pipe.
Therefore we endow them with the cross section of their respective attached pipes.
We do so in order to have the coupling reach through the entire network.
Otherwise, there would be three distinct parts, one before the compressor, one after the valve and one in between, that are not coupled through the Bernoulli coupling.
The inflow of gas into the three source nodes and outflow at (non-gas plant) sink nodes of the network is chosen constant.
A list of both can be found in\begin{table}
  \centering
  \caption{Volumetric inflow and outflow at source and sink nodes of the gas network.}
  \label{tab:in-and-outflow}
  \begin{subtable}{0.49\textwidth}
    \centering
    \caption{Volumetric inflow for gas network.}
    \begin{tabular}[ht]{lr}
      \toprule
      Node ID & Inflow $[ \si{ \cubic\metre \per \second} ]$\\
      \midrule
      node\textunderscore1 & 58.993631 \\
      node\textunderscore20&190.815287 \\
      node\textunderscore80& 61.866242 \\
      \bottomrule
    \end{tabular}
    \label{tab:Inflow-gas}
  \end{subtable}
  \begin{subtable}{0.49\textwidth}
    \centering
    \caption{Volumetric outflow at sinks other than conversion plants.}
      \begin{tabular}[ht]{lr}
    \toprule
    Gas network ID & Outflow $[ \si{ \cubic\metre \per \second} ]$\\
    \midrule
    node\textunderscore ld1  &$0.000000$\\
    node\textunderscore ld3  &$0.000000$\\
    node\textunderscore ld4  &$0.121019$\\
    node\textunderscore ld5  &$0.000000$\\
    node\textunderscore ld7  &$1.490446$\\
    node\textunderscore ld8  &$2.089172$\\
    node\textunderscore ld9  &$0.000000$\\
    node\textunderscore ld11 &$5.490446$\\
    node\textunderscore ld14 &$0.452229$\\
    node\textunderscore ld15 &$0.280255$\\
    node\textunderscore ld16 &$0.076433$\\
    node\textunderscore ld17 &$4.617834$\\
    node\textunderscore ld18 &$4.617834$\\
    node\textunderscore ld19 &$0.802548$\\
    node\textunderscore ld20 &$0.445860$\\
    node\textunderscore ld21 &$0.286624$\\
    node\textunderscore ld22 &$7.592357$\\
    node\textunderscore ld23 &$0.082803$\\
    node\textunderscore ld25 &$0.802548$\\
    node\textunderscore ld26 &$0.000000$\\
    node\textunderscore ld27 &$0.012739$\\
    node\textunderscore ld28 &$0.000000$\\
    node\textunderscore ld30 &$1.426752$\\
    node\textunderscore ld32 &$0.000000$\\
    node\textunderscore ld33 &$1.101911$\\
    node\textunderscore ld34 &$0.000000$\\
    node\textunderscore ld35 &$0.000000$\\
    node\textunderscore ld37 &$7.732484$\\
    node\textunderscore ld38 &$0.000000$\\
    node\textunderscore ld39 &$0.000000$\\
    node\textunderscore ld40 &$7.732484$\\
    node\textunderscore ld41 &$1.528662$\\
    node\textunderscore ld43 &$0.000000$\\
    node\textunderscore ld44 &$0.000000$\\
    node\textunderscore ld45 &$0.000000$\\
    \bottomrule
  \end{tabular}


    \label{tab:other-outflow}
  \end{subtable}
  \end{table}

\subsection{Power network}
\label{sec:power-network}
For the power model we adapt the IEEE 300-bus test case that is part of the Matpower software~\cite{Zimmerman11}.
Originally, this system has a total of $\nbus = 300$ buses (1 slack bus, 68 \PVbus buses, 231 \PQbus buses), and $\nline =  411$ lines.
We modify the grid such that the original slack bus is now a \PVbus bus, and the nodes listed in \autoref{tab:gas-power-interface} are all slack buses.
These are linked to sinks of the gas network.
At these buses, gas and electricity can be converted into each other here.
The IDs of the connected sinks in the gas network are given in the table.
\begin{table}
  \centering
  \caption{Connection of gas nodes and power nodes for conversion.}
  \begin{tabular}{rr}
    \toprule
    Power grid ID & Gas network ID \\
    \midrule
    213  &node\textunderscore ld31\\
    221  &node\textunderscore ld24\\
    230  &node\textunderscore ld13\\
    7001 &node\textunderscore ld36\\
    7017 &node\textunderscore ld2\phantom{3}\\
    7024 &node\textunderscore ld12\\
    7039 &node\textunderscore ld42\\
    7057 &node\textunderscore ld6\phantom{3}\\
    7061 &node\textunderscore ld29\\
    7071 &node\textunderscore ld10\\
    \bottomrule
  \end{tabular}
  \label{tab:gas-power-interface}
\end{table}
Therefore we have a total of $10$ slack buses, $59$ \PVbus buses and $231$ \PQbus buses.
The nominal total active power generation of the grid is about $24,000\ \si{MW}$.
For the slack buses and the \PVbus buses, we use the bus specifications from the original case file.
For \PQbus buses we use time-dependent bus specifications given by
\begin{equation}
  \label{eq:6}
  \begin{aligned}
    \activepower(t) = \activepower_{300} \left( 0.9+0.4 \sin \left(\frac{2\pi t}{24 \si{\hour}}\right)\right),\\
    \reactivepower(t) = \reactivepower_{300} \left(0.9+0.4 \sin \left(\frac{2\pi t}{24 \si{\hour}}\right)\right),
\end{aligned}
\end{equation}
where $\activepower_{300}$ and $\reactivepower_{300}$ are the active and reactive power demand from the original case file.

\subsubsection{Parameters of Gas-Power conversion}
\label{sec:parameters-gas-power}
We need to specify values for the conversion factors $E_\text{GtP}$ and $E_\text{PtG}$ in~\eqref{eq:4}.For the operation as a gas power plant we choose an efficiency of $\eta_{\text{GtP}} = 0.4$ with respect to the lower heating value of the gas.
This is a realistic value, given that there are gas power plants with efficiencies of up to $60 \, \%$ \cite{ens.dk19}. The lower heating value $L$ of natural gas is usually in the range of $36 \si{\mega\J \per \kilogram}\leq L \leq 50 \si{\mega\J \per \kilogram}$, depending on the gas composition~\cite{cerbe2016grundlagen}.
We choose $L =40 \si{\mega\J \per \kilogram}$.
The parameter $E_\text{GtP}$ is then obtained from
\begin{equation}
  \label{eq:14}
  E_\text{GtP} = \frac{1}{\rho_0 L \eta_{\text{GtP}}} \approx \SI{0.0796}{\cubic\metre \per \mega \J}.
\end{equation}
For PtG conversion we choose an efficiency of $\eta_\text{PtG} = 0.8$, this time with respect to the upper heating value according to \cite{pothochtempelyse}. The upper heating value $U$ of natural gas is given by $U = 1.11 L$.
Therefore we obtain
\begin{equation}
\label{eq:21}
E_\text{PtG} = \frac{\eta_{\text{PtG}}}{\rho_0  U} \approx \SI{0.0229}{\cubic\metre \per \mega \J}.
\end{equation}

\section{Numerical results}
Using the network data from \autoref{sec:network-data} we simulate the combined network over a time horizon of 24 hours.
Each simulation run took about $20$ seconds on an \emph{Intel(R) Core(TM) i7-7700 CPU @ 3.60GHz} with $32$ Gigabytes of RAM.
\subsection{Comparison of coupling conditions}
\label{sec:comp-coupl-cond}
During implementation of the Bernoulli coupling constraint it became clear that the choice of coupling condition $H_b$ over $H_p$ introduced almost no difference.To quantify the difference we simulate once (simulation ``p'') with the pressure coupling constant $H_p$, and once (simulation ``b'') with the Bernoulli coupling constant $H_b$.
We compare the values of the pressures and the volumetric flows on the whole gas network at every timestep.
Let $p_p(t,x)$ be the pressure obtained from simulation ``p'' at time $t$ and at some position in the network ($x$ ranges over all pipes and all pipe lengths).Further, let $p_b(t,x)$ be the analogue for simulation ``b'' and let $q_p(t,x)$ and $q_b(t,x)$ the corresponding values for the volumetric flow.
\autoref{tab:coupling-results} shows our findings with regard to the different coupling constants.
\begin{table}
  \centering
  \caption{Absolute and relative differences for the two coupling constants.}
  \label{tab:coupling-results}
   \newlength{\oldextrarowheight}
  \setlength{\oldextrarowheight}{\extrarowheight}
  \setlength{\extrarowheight}{3pt}
  \newlength{\oldtabcolsep}
  \setlength{\oldtabcolsep}{\tabcolsep}
  \setlength{\tabcolsep}{5pt}
  \begin{subtable}[ht]{1\textwidth}
    \centering
    \caption{Absolute and relative difference of pressure.}
      \label{tab:simple-pressure-results}
    \begin{tabular}[ht]{rr}
      \toprule
      $\max{\abs{p_p-p_b}}$ &$\max{\tfrac{\abs{p_p-p_b}}{p_p}}$\\
      \midrule
      \SI{0.1828}{\bar}&$0.0041$\\
      \bottomrule      
    \end{tabular}
      \end{subtable}
  
  \bigskip
  \begin{subtable}[ht]{1\textwidth}
    \centering
    \caption{Absolute and relative difference of flow for different cut-off values.}
    \label{tab:relative-q-results}
    \begin{tabular}[ht]{lrr}
      \toprule
      range [\si{\cubic\metre\per\second}]&$\max\abs{q_p-q_b}$&$\max\tfrac{\abs{q_p-q_b}}{\abs{q_p}}$\\
      \midrule
      $10^{-3}<\abs{q_p}<10^{-2}$&\SI{0.0004}{\cubic\metre\per\second}&0.3270\\
      $10^{-2}<\abs{q_p}<10^{-1}$&\SI{0.0180}{\cubic\metre\per\second}&0.32670\\
      $10^{-1}<\abs{q_p}<10^{0\phantom{{}-{}}}$&\SI{0.0225}{\cubic\metre\per\second}&0.1105\\
      $10^{0\phantom{{}-{}}}<\abs{q_p}<10^{1\phantom{{}-{}}}$&\SI{0.0227}{\cubic\metre\per\second}&0.0207\\
      $10^{1\phantom{{}-{}}}<\abs{q_p}\phantom{{}<10^{-2}}$&\SI{0.0570}{\cubic\metre\per\second}& 0.0029\\
      \bottomrule                                                              
    \end{tabular}
  \end{subtable}
  \setlength{\tabcolsep}{\oldtabcolsep}
  \setlength{\extrarowheight}{\oldextrarowheight}


 \end{table}
 For the relative differences in the flow we used different cut-off values, because although the relative error grows when approaching $q = 0$, the absolute values are very small and hence probably of little significance.In contrast to~\cite{reigstad2} we find little difference for the two coupling conditions.
 The key difference is the absence of a friction term in~\cite{reigstad2}, which allows errors to accumulate.
 In our case artificial energy produced at the nodes is consumed by friction and cannot cause much error. 
 In light of the small size of the error introduced by using the physically unsound pressure coupling constant practitioners should trade-off carefully the need for more accuracy against the practical hurdles mentioned in~\autoref{remark:nodes}.

\subsection{Gas-power-conversion}
\label{sec:gas-power-conv}
\begin{figure}[!ht]
  \centering
  \includegraphics[width=\textwidth]{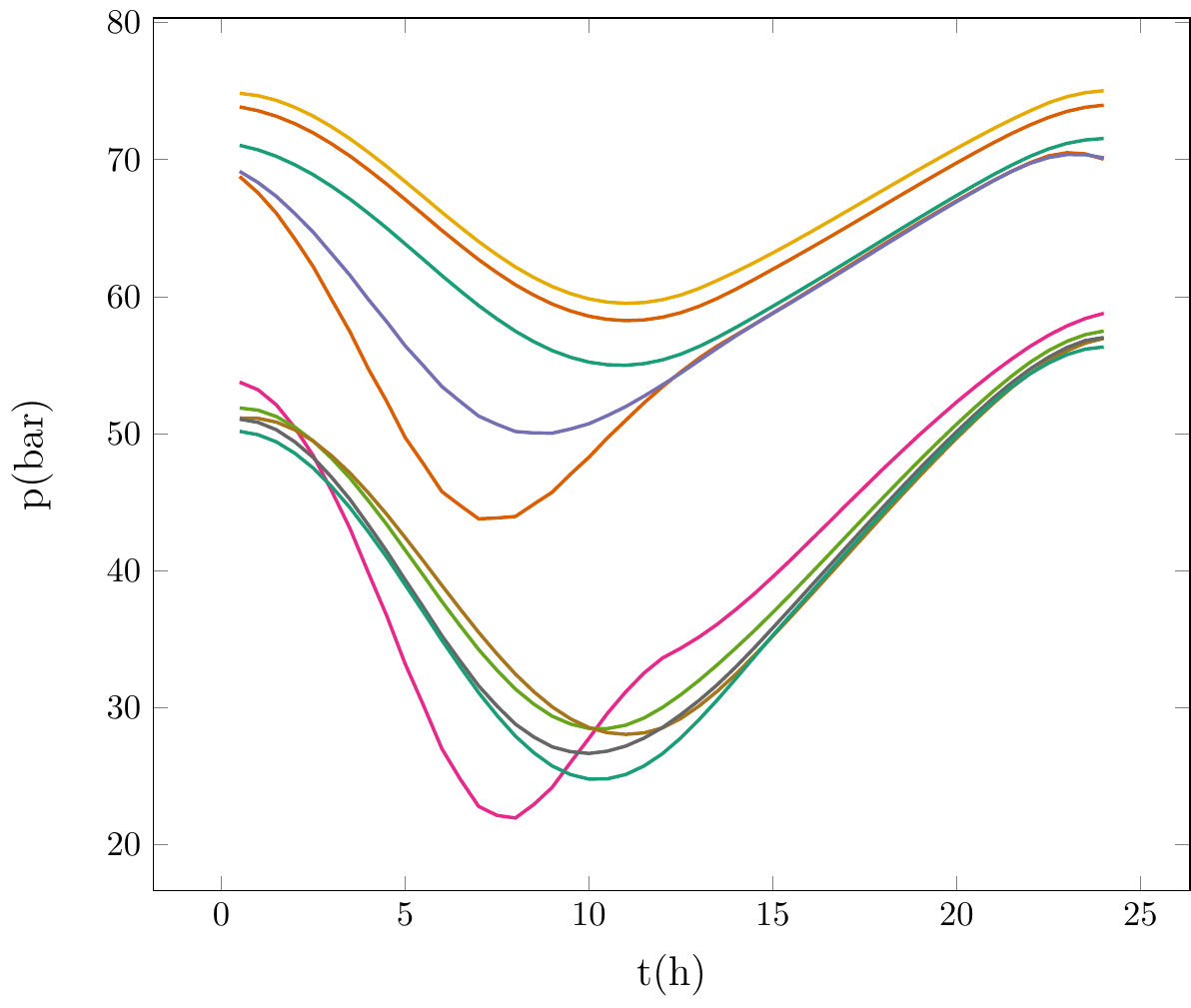}
  \caption{Pressure at the gas-plant nodes over time.}
  \label{fig:pressure}
\end{figure}
We now present the results of Gas-to-Power conversion and Power-to-Gas conversion.
Over a day all of the plants go through a cycle of high power demand during which gas is consumed to power a generator.
During the second half of the day much less power is needed and so the Gas-to-Power mode is used to convert power back to gas.
All the data of pressure and flow in the conversion plants is found in \autoref{tab:pressure-gtp} and \autoref{tab:flow-gtp} on pages \pageref{tab:pressure-gtp} and \pageref{tab:flow-gtp}.
The total volume of gas consumed by the power plants is obtained by integrating the outflow over time: using the trapezoidal rule, in our case it is \SI{2.3098e+07}{\cubic\metre}, the total volume of gas generated is \SI{2.0522e+06}{\cubic\metre}.
\begin{figure}[!ht]
  \centering
  \includegraphics[width=\textwidth]{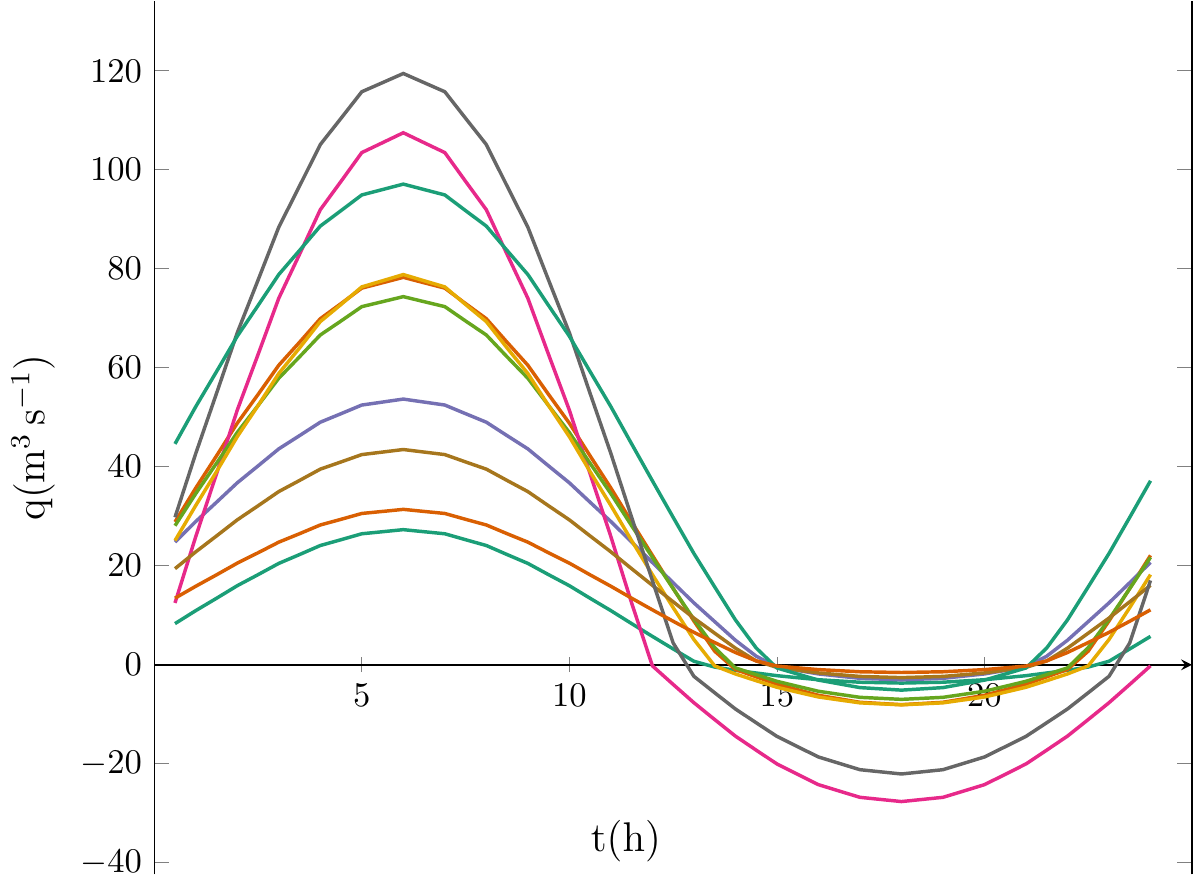}
  \caption{Gas flow for power conversion over time.}
  \label{fig:flow}
\end{figure}
\autoref{fig:pressure} shows the pressure evolution at the conversion nodes.
It shows that the gas network cannot provide the peak power demand indefinitely as the pressure drops considerably during power generation (in the first 12 hours).
But it is suitable to counter balance high and low power demand over the course of a day as it recuperates during low power demand when gas is injected into the pipeline network by power-to-gas operation.
\label{sec:numerical-results}

\autoref{fig:flow} shows the amount of gas consumed ($q >0$) by power generation and generated ($q<0$) by power-to-gas operation.
Note the kink at $q=0$ which is due to the difference in efficiencies of the two processes.

\section{Conclusion}
\label{sec:conclusion}
We have presented a model for a combined gas and power network and simulated the operation of it over a time horizon of $24$ hours.
We evaluated the impact of two different coupling conditions, namely pressure coupling and Bernoulli coupling, and found them to be negligible for practical purposes of pipeline simulation.
In addition our simulation results showed the given gas network to be able to provide enough power to counter balance power demand peaks and our tools provide visualization and quantification of gas consumed and produced.
Our data can be used to benchmark similar tools.  The data can be found under \url{https://bitbucket.org/efokken/gas-power-benchmark/src/master/}.

Future work includes the solution of corresponding optimal control problems and the inclusion of stochastic effects.

\printbibliography

\begin{table}[!h]
  \centering
  \caption{Pressure values at the gas-power conversion plants (time in hours,
    pressure in \si{\bar}).}
  \label{tab:pressure-gtp}
\begin {tabular}{S[round-mode=places,round-precision=1]S[table-format=3.2,round-mode=places,round-precision=2]S[table-format=3.2,round-mode=places,round-precision=2]S[table-format=3.2,round-mode=places,round-precision=2]S[table-format=3.2,round-mode=places,round-precision=2]S[table-format=3.2,round-mode=places,round-precision=2]S[table-format=3.2,round-mode=places,round-precision=2]S[table-format=3.2,round-mode=places,round-precision=2]S[table-format=3.2,round-mode=places,round-precision=2]S[table-format=3.2,round-mode=places,round-precision=2]S[table-format=3.2,round-mode=places,round-precision=2]}%
\toprule \multicolumn {1}{c}{t}&\multicolumn {1}{c}{N7071}&\multicolumn {1}{c}{N7024}&\multicolumn {1}{c}{N230}&\multicolumn {1}{c}{N221}&\multicolumn {1}{c}{N7061}&\multicolumn {1}{c}{N7017}&\multicolumn {1}{c}{N213}&\multicolumn {1}{c}{N7001}&\multicolumn {1}{c}{N7039}&\multicolumn {1}{c}{N7057}\\\midrule %
\rowcolor [gray]{0.85}0.5000&71.0455&68.7749&69.1475&53.7685&51.8813&74.8464&51.1480&51.0639&50.1773&73.8533\\%
1.0000&70.7183&67.5806&68.3356&53.2087&51.7221&74.6727&51.1268&50.8318&49.9303&73.5722\\%
\rowcolor [gray]{0.85}1.5000&70.2439&66.0861&67.3185&52.1056&51.2539&74.3291&50.8509&50.2887&49.4061&73.1693\\%
2.0000&69.6307&64.2225&66.0704&50.4439&50.4842&73.8227&50.2854&49.4100&48.5786&72.6254\\%
\rowcolor [gray]{0.85}2.5000&68.9075&62.2002&64.7148&48.4085&49.4746&73.1821&49.4661&48.2694&47.5002&71.9598\\%
3.0000&68.0585&59.8037&63.1425&45.8429&48.2067&72.4049&48.3963&46.8280&46.1479&71.1605\\%
\rowcolor [gray]{0.85}3.5000&67.1256&57.4434&61.5661&43.0930&46.7655&71.5286&47.1358&45.2112&44.6087&70.2682\\%
4.0000&66.0892&54.7383&59.8057&39.8636&45.1188&70.5470&45.6883&43.3498&42.8498&69.2680\\%
\rowcolor [gray]{0.85}4.5000&65.0069&52.3600&58.1767&36.7211&43.3824&69.5084&44.1217&41.4267&40.9891&68.2175\\%
5.0000&63.8572&49.7330&56.4249&33.2189&41.5144&68.4044&42.4360&39.3420&38.9853&67.0974\\%
\rowcolor [gray]{0.85}5.5000&62.7105&47.8000&54.9567&30.1795&39.6577&67.2927&40.7078&37.3300&36.9822&65.9799\\%
6.0000&61.5441&45.7813&53.4489&27.0182&37.7607&66.1634&38.9373&35.2621&34.9319&64.8411\\%
\rowcolor [gray]{0.85}6.5000&60.4340&44.7646&52.3513&24.7939&35.9811&65.0778&37.2053&33.4050&32.9930&63.7592\\%
7.0000&59.3556&43.7928&51.2905&22.8096&34.2578&64.0244&35.5113&31.6069&31.1123&62.7058\\%
\rowcolor [gray]{0.85}7.5000&58.3800&43.8553&50.6880&22.1493&32.7428&63.0595&33.9318&30.1305&29.4455&61.7526\\%
8.0000&57.4819&43.9650&50.1644&21.9610&31.3714&62.1701&32.4653&28.8127&27.9376&60.8702\\%
\rowcolor [gray]{0.85}8.5000&56.7205&44.8620&50.0599&22.9462&30.2741&61.4023&31.1786&27.8732&26.7254&60.1165\\%
9.0000&56.0720&45.7336&50.0430&24.1956&29.3906&60.7433&30.0709&27.1528&25.7559&59.4658\\%
\rowcolor [gray]{0.85}9.5000&55.5805&47.0555&50.3491&26.0187&28.8166&60.2267&29.1967&26.7986&25.1315&58.9597\\%
10.0000&55.2262&48.2887&50.7319&27.7734&28.4977&59.8413&28.5574&26.6718&24.8009&58.5786\\%
\rowcolor [gray]{0.85}10.5000&55.0366&49.6904&51.3252&29.5911&28.4850&59.6077&28.1880&26.8397&24.8182&58.3477\\%
11.0000&54.9968&50.9792&51.9798&31.1656&28.7349&59.5170&28.0627&27.2097&25.1356&58.2543\\%
\rowcolor [gray]{0.85}11.5000&55.1181&52.2577&52.7481&32.5462&29.2599&59.5771&28.1738&27.7955&25.7606&58.3085\\%
12.0000&55.3914&53.4322&53.5678&33.6339&30.0288&59.7817&28.5369&28.5618&26.6578&58.5035\\%
\rowcolor [gray]{0.85}12.5000&55.8117&54.5168&54.4360&34.3566&30.9620&60.1269&29.2231&29.5118&27.8101&58.8370\\%
13.0000&56.3744&55.5234&55.3512&35.1781&32.0166&60.6090&30.1445&30.5516&29.1532&59.3059\\%
\rowcolor [gray]{0.85}13.5000&57.0377&56.4177&56.2633&36.1206&33.1627&61.2019&31.2350&31.7098&30.6045&59.8902\\%
14.0000&57.7644&57.2086&57.1477&37.1864&34.3806&61.8273&32.4753&33.0018&32.1458&60.5551\\%
\rowcolor [gray]{0.85}14.5000&58.5205&57.9972&57.9750&38.3290&35.6358&62.4940&33.8232&34.3851&33.7216&61.2679\\%
15.0000&59.3027&58.8135&58.7796&39.5603&36.9585&63.1991&35.2322&35.8317&35.2923&62.0055\\%
\rowcolor [gray]{0.85}15.5000&60.0868&59.6250&59.5729&40.8282&38.3172&63.9286&36.6674&37.2919&36.8079&62.7516\\%
16.0000&60.8857&60.4567&60.3846&42.1504&39.7089&64.6803&38.1296&38.7701&38.3199&63.5143\\%
\rowcolor [gray]{0.85}16.5000&61.6906&61.2860&61.2007&43.4672&41.1107&65.4437&39.6022&40.2453&39.8220&64.2869\\%
17.0000&62.5088&62.1313&62.0320&44.8103&42.5271&66.2192&41.0866&41.7276&41.3251&65.0711\\%
\rowcolor [gray]{0.85}17.5000&63.3278&62.9626&62.8593&46.1200&43.9358&66.9971&42.5654&43.1923&42.8105&65.8588\\%
18.0000&64.1539&63.8019&63.6946&47.4359&45.3425&67.7787&44.0400&44.6500&44.2854&66.6514\\%
\rowcolor [gray]{0.85}18.5000&64.9731&64.6178&64.5172&48.7006&46.7247&68.5546&45.4923&46.0749&45.7290&67.4402\\%
19.0000&65.7912&65.4336&65.3394&49.9551&48.0882&69.3264&46.9234&47.4773&47.1472&68.2263\\%
\rowcolor [gray]{0.85}19.5000&66.5942&66.2208&66.1419&51.1469&49.4114&70.0848&48.3157&48.8328&48.5196&69.0008\\%
20.0000&67.3871&67.0006&66.9357&52.3142&50.6996&70.8313&49.6701&50.1506&49.8515&69.7646\\%
\rowcolor [gray]{0.85}20.5000&68.1570&67.7502&67.7045&53.4132&51.9333&71.5570&50.9705&51.4093&51.1240&70.5093\\%
21.0000&68.9080&68.4857&68.4565&54.4761&53.1169&72.2629&52.2177&52.6164&52.3416&71.2351\\%
\rowcolor [gray]{0.85}21.5000&69.6079&69.1644&69.1454&55.4625&54.2182&72.9363&53.3731&53.7289&53.4334&71.9165\\%
22.0000&70.2464&69.7832&69.7330&56.3825&55.2219&73.5701&54.4081&54.7299&54.3815&72.5376\\%
\rowcolor [gray]{0.85}22.5000&70.7852&70.2768&70.1534&57.1860&56.0680&74.1509&55.3028&55.5863&55.1608&73.0851\\%
23.0000&71.1909&70.5080&70.3734&57.8705&56.7468&74.5951&56.0517&56.2940&55.7739&73.5244\\%
\rowcolor [gray]{0.85}23.5000&71.4407&70.4238&70.3619&58.4119&57.2346&74.8895&56.6179&56.8014&56.1775&73.8243\\%
24.0000&71.5473&70.0346&70.1380&58.7883&57.4952&75.0361&56.9596&57.0237&56.3332&73.9787\\\bottomrule %
\end {tabular}%

\end{table}

\begin{table}[!h]
  \centering
  \caption{Flow values at the gas-power conversion plants (time in hours,
    in \si{\cubic\metre\per\second}). Positive flow means gas is converted to power,
    negative flow means gas is generated with power.}
  \label{tab:flow-gtp}
\begin {tabular}{S[round-mode=places,round-precision=1]S[round-mode=places,round-precision=2]S[round-mode=places,round-precision=2]S[round-mode=places,round-precision=2]S[round-mode=places,round-precision=2]S[round-mode=places,round-precision=2]S[round-mode=places,round-precision=2]S[round-mode=places,round-precision=2]S[round-mode=places,round-precision=2]S[round-mode=places,round-precision=2]S[round-mode=places,round-precision=2]}%
\toprule \multicolumn {1}{c}{t}&\multicolumn {1}{c}{N7071}&\multicolumn {1}{c}{N7024}&\multicolumn {1}{c}{N230}&\multicolumn {1}{c}{N221}&\multicolumn {1}{c}{N7061}&\multicolumn {1}{c}{N7017}&\multicolumn {1}{c}{N213}&\multicolumn {1}{c}{N7001}&\multicolumn {1}{c}{N7039}&\multicolumn {1}{c}{N7057}\\\midrule %
\rowcolor [gray]{0.85}0.5000&8.3239&28.8649&24.8018&12.4990&28.1027&25.0951&19.4250&29.8060&44.6346&13.4619\\%
1.0000&10.9487&35.7187&28.9484&25.9047&34.5863&32.1913&22.8247&42.6988&52.1582&15.8722\\%
\rowcolor [gray]{0.85}1.5000&13.4409&42.2100&32.8427&38.5824&40.7160&39.0234&26.0325&54.8326&59.2368&18.1753\\%
2.0000&15.9830&48.8202&36.7720&51.4521&46.9399&46.0882&29.2849&67.0847&66.3854&20.5409\\%
\rowcolor [gray]{0.85}2.5000&18.2112&54.6092&40.1799&62.6711&52.3712&52.3627&32.1192&77.7048&72.5857&22.6303\\%
3.0000&20.4881&60.5241&43.6260&74.0623&57.8987&58.8562&34.9989&88.4243&78.8518&24.7838\\%
\rowcolor [gray]{0.85}3.5000&22.2743&65.1663&46.3020&82.9349&62.2208&64.0105&37.2452&96.7253&83.7131&26.4890\\%
4.0000&24.1016&69.9199&49.0121&91.9391&66.6334&69.3407&39.5298&105.1035&88.6320&28.2512\\%
\rowcolor [gray]{0.85}4.5000&25.2758&72.9781&50.7371&97.6769&69.4672&72.7977&40.9895&110.4176&91.7613&29.3956\\%
5.0000&26.4741&76.1035&52.4825&103.4846&72.3621&76.3526&42.4712&115.7769&94.9279&30.5756\\%
\rowcolor [gray]{0.85}5.5000&26.8893&77.1876&53.0833&105.4835&73.3669&77.5910&42.9824&117.6172&96.0183&30.9879\\%
6.0000&27.3082&78.2823&53.6873&107.4927&74.3824&78.8442&43.4970&119.4648&97.1151&31.4061\\%
\rowcolor [gray]{0.85}6.5000&26.8893&77.1876&53.0833&105.4835&73.3669&77.5910&42.9824&117.6172&96.0183&30.9879\\%
7.0000&26.4741&76.1035&52.4825&103.4846&72.3621&76.3526&42.4712&115.7769&94.9279&30.5756\\%
\rowcolor [gray]{0.85}7.5000&25.2758&72.9781&50.7371&97.6769&69.4672&72.7977&40.9895&110.4176&91.7613&29.3956\\%
8.0000&24.1016&69.9199&49.0121&91.9391&66.6334&69.3407&39.5298&105.1035&88.6320&28.2512\\%
\rowcolor [gray]{0.85}8.5000&22.2743&65.1663&46.3020&82.9349&62.2208&64.0105&37.2452&96.7253&83.7131&26.4890\\%
9.0000&20.4881&60.5241&43.6260&74.0623&57.8987&58.8562&34.9989&88.4243&78.8518&24.7838\\%
\rowcolor [gray]{0.85}9.5000&18.2112&54.6092&40.1799&62.6711&52.3712&52.3627&32.1192&77.7048&72.5857&22.6303\\%
10.0000&15.9830&48.8202&36.7720&51.4521&46.9399&46.0882&29.2849&67.0847&66.3854&20.5409\\%
\rowcolor [gray]{0.85}10.5000&13.4409&42.2100&32.8427&38.5824&40.7160&39.0234&26.0325&54.8326&59.2368&18.1753\\%
11.0000&10.9487&35.7187&28.9484&25.9047&34.5863&32.1913&22.8247&42.6988&52.1582&15.8722\\%
\rowcolor [gray]{0.85}11.5000&8.3239&28.8649&24.8018&12.4990&28.1027&25.0951&19.4250&29.8060&44.6346&13.4619\\%
12.0000&5.7461&22.1108&20.6834&-0.2049&21.7107&18.2245&16.0640&17.0461&37.1834&11.1087\\%
\rowcolor [gray]{0.85}12.5000&3.2120&15.4428&16.5889&-3.9583&15.4078&11.5666&12.7370&4.4213&29.8041&8.8081\\%
13.0000&0.7195&8.8495&12.5149&-7.6573&9.1946&5.1122&9.4404&-2.3249&22.4972&6.5572\\%
\rowcolor [gray]{0.85}13.5000&-0.4518&2.7640&8.7345&-11.0546&3.4888&-0.2088&6.3931&-5.6415&15.7555&4.5029\\%
14.0000&-1.1012&-0.9434&4.9679&-14.4022&-0.6141&-1.8424&3.3676&-8.9201&9.0817&2.4899\\%
\rowcolor [gray]{0.85}14.5000&-1.6511&-2.4281&1.7433&-17.2346&-1.9831&-3.2062&0.7856&-11.7036&3.4081&0.7944\\%
15.0000&-2.1938&-3.9052&-0.4246&-20.0245&-3.3294&-4.5335&-0.5138&-14.4560&-0.6368&-0.2506\\%
\rowcolor [gray]{0.85}15.5000&-2.6052&-5.0343&-1.1344&-22.1333&-4.3450&-5.5267&-1.0792&-16.5457&-1.8675&-0.6122\\%
16.0000&-3.0121&-6.1605&-1.8428&-24.2108&-5.3433&-6.4974&-1.6424&-18.6152&-3.0871&-0.9677\\%
\rowcolor [gray]{0.85}16.5000&-3.2655&-6.8672&-2.2874&-25.4986&-5.9609&-7.0957&-1.9952&-19.9052&-3.8476&-1.1879\\%
17.0000&-3.5170&-7.5734&-2.7314&-26.7704&-6.5697&-7.6846&-2.3472&-21.1865&-4.6031&-1.4055\\%
\rowcolor [gray]{0.85}17.5000&-3.6024&-7.8142&-2.8826&-27.2001&-6.7753&-7.8832&-2.4670&-21.6215&-4.8595&-1.4790\\%
18.0000&-3.6875&-8.0549&-3.0338&-27.6276&-6.9796&-8.0807&-2.5867&-22.0555&-5.1153&-1.5522\\%
\rowcolor [gray]{0.85}18.5000&-3.6024&-7.8142&-2.8826&-27.2001&-6.7753&-7.8832&-2.4670&-21.6215&-4.8595&-1.4790\\%
19.0000&-3.5170&-7.5734&-2.7314&-26.7704&-6.5697&-7.6846&-2.3472&-21.1865&-4.6031&-1.4055\\%
\rowcolor [gray]{0.85}19.5000&-3.2655&-6.8672&-2.2874&-25.4986&-5.9609&-7.0957&-1.9952&-19.9052&-3.8476&-1.1879\\%
20.0000&-3.0121&-6.1605&-1.8428&-24.2108&-5.3433&-6.4974&-1.6424&-18.6152&-3.0871&-0.9677\\%
\rowcolor [gray]{0.85}20.5000&-2.6052&-5.0343&-1.1344&-22.1333&-4.3450&-5.5267&-1.0792&-16.5457&-1.8675&-0.6122\\%
21.0000&-2.1938&-3.9052&-0.4246&-20.0245&-3.3294&-4.5335&-0.5138&-14.4560&-0.6368&-0.2506\\%
\rowcolor [gray]{0.85}21.5000&-1.6511&-2.4281&1.7433&-17.2346&-1.9831&-3.2062&0.7856&-11.7036&3.4081&0.7944\\%
22.0000&-1.1012&-0.9434&4.9679&-14.4022&-0.6141&-1.8424&3.3676&-8.9201&9.0817&2.4899\\%
\rowcolor [gray]{0.85}22.5000&-0.4518&2.7640&8.7345&-11.0546&3.4888&-0.2088&6.3931&-5.6415&15.7555&4.5029\\%
23.0000&0.7195&8.8495&12.5149&-7.6573&9.1946&5.1122&9.4404&-2.3249&22.4972&6.5572\\%
\rowcolor [gray]{0.85}23.5000&3.2120&15.4428&16.5889&-3.9583&15.4078&11.5666&12.7370&4.4213&29.8041&8.8081\\%
24.0000&5.7461&22.1108&20.6834&-0.2049&21.7107&18.2245&16.0640&17.0461&37.1834&11.1087\\\bottomrule %
\end {tabular}%

\end{table}

\end{document}